\documentclass[%
 reprint,
superscriptaddress,
 amsmath,amssymb,
 aps,
prapplied, floatfix, longbibliography ]{revtex4-2}
\usepackage{graphicx}
\usepackage{dcolumn}
\usepackage{bm}
\usepackage[colorlinks,linkcolor=blue,anchorcolor=blue,citecolor=blue,urlcolor=blue]{hyperref}
\begin{document}

\title{Reversable phase transitions in ferroic two-dimensional
       Nb$_2$O$_2$I$_4$ through optically excited coherent phonons}

\author{Chuanlin~Liu}
\affiliation{School of Physics,
             Southeast University,
             Nanjing, 211189, PRC}

\author{Dan~Liu}
\email      {liudan2@seu.edu.cn}%
\affiliation{School of Physics,
             Southeast University,
             Nanjing, 211189, PRC}
\affiliation{Key Laboratory of Quantum Materials and Devices of Ministry of Education,
             Southeast University,
             Nanjing, 211189, PRC}

\author{Jie~Guan}
\affiliation{School of Physics,
             Southeast University,
             Nanjing, 211189, PRC}

\author{Chao~Lian}
\email      {chaolian@iphy.ac.cn}%
\affiliation{Beijing National Laboratory for Condensed Matter Physics and Institute of Physics,
             Chinese Academy of Sciences,
             Beijing, 100190, P. R. China}
\affiliation{Songshan Lake Materials Laboratory,
             Dongguan City, Guangdong Province, P. R. China}

\date{\today}

\begin{abstract}
We investigate optically induced phase transitions in the
two-dimensional (2D) ferroelectric (FE) material Nb$_2$O$_2$I$_4$
using real-time time-dependent density functional theory
(rt-TDDFT). Our results demonstrate that tailored laser pulses can
activate specific coherent phonon modes. Specifically, the
anharmonic atomic distortions of the A1-1 and A1-2 modes at the
$\Gamma$-point facilitate the reversal of in-plane polarization. By
fine-tuning laser parameters, additional phonon modes at both the
$Y$ and $\Gamma$ points are excited. The resulting nonequilibrium
atomic dynamics enable the formation of previously unreported
ferroic phases, including three antiferroelectric (AFE) phases and
one ferrielectric (FiE) phase. Notably, these optically induced
phases can be reverted to the initial FE state using appropriate
techniques. This controllable reversibility among multiple ferroic
phases positions 2D Nb$_2$O$_2$I$_4$ as a highly promising
candidate for next-generation electronic storage applications.
\end{abstract}




\maketitle
\renewcommand\thesubsection{\arabic{subsection}}




\section{Introduction}

The escalating demand for data storage in the electronics industry
has intensified the search for materials and technologies that
offer superior power efficiency, rapid write speeds, high
endurance, and increased storage
density~\cite{{1},{2},{3},{4},{5},{6},{7},{7-1},{7-2}}. While
conventional solutions like hard disk drives and flash memory
remain dominant, emerging alternatives such as ferroelectric RAM
(F-RAM) are gaining traction. F-RAM offers distinct advantages,
including lower power consumption, faster write speeds, and
read/write endurance that significantly surpasses that of flash
memory~\cite{{reddy},{9},{10}}. However, F-RAM also has
limitations, such as lower storage densities and response times on
the order of nanoseconds. These drawbacks primarily stem from the
mechanism of flipping ferroelectric polarization using conventional
electric fields~\cite{{11},{12}}. To address these limitations,
femtosecond lasers offer a promising alternative for manipulating
ferroelectric states, potentially improving response speeds from
nanoseconds to picoseconds and enhancing storage
density~\cite{{13J},{14Q},{gu24},{16Q},{19cj},{20-1},{20-2},{20-3},{20-4}}.
Realizing this potential requires precise optical control over
ferroelectric polarization states. Furthermore, if optically
induced polarization changes beyond simple polarization reversal,
such materials could be used to support non-binary storage,
allowing more data to be stored per cell than in traditional memory
technologies~\cite{{17},{18},{17-1},{17-2}}.

\begin{figure*}[t]
\includegraphics[width=1.8\columnwidth]{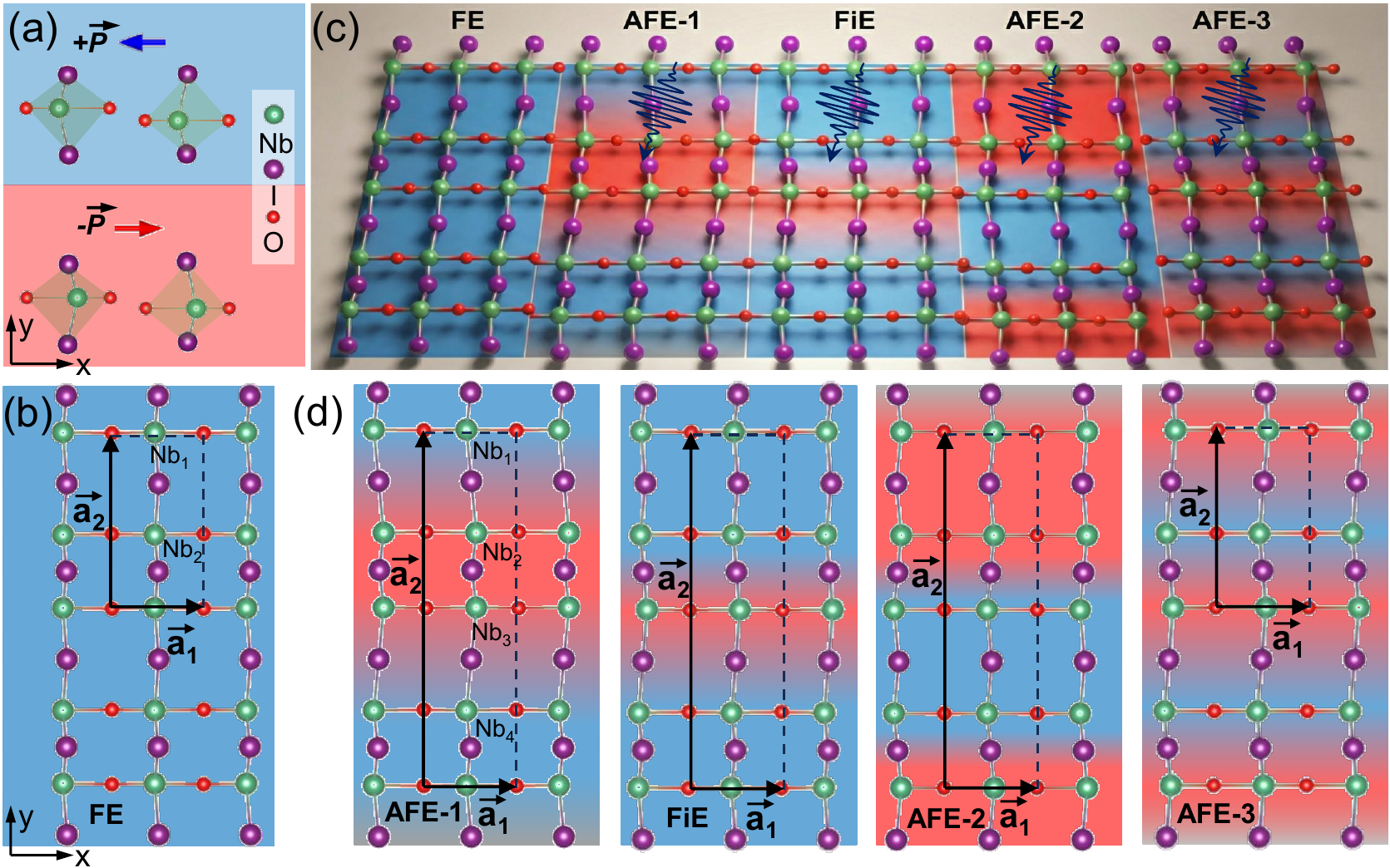}
\caption{%
(a) The local electrical polarization of the Nb-O-I cage, the blue
and red arrows indicate the polarization pointing to the left and
right direction. (b) The atomic structure of the FE phase of
Nb$_2$O$_2$I$_4$. (c) Schematic illustration of phase
transformation from the 2D FE-Nb$_2$O$_2$I$_4$ to the other ferroic
phases with illumination of laser pulses. (d) The atomic structure
of three AFE phases and one FiE phase of Nb$_2$O$_2$I$_4$. The
background with blue and red color represents local electrical
polarization of the Nb-O-I cages at this area with direction
pointing to left and right.%
\label{fig1}}
\end{figure*}


Utilizing time-dependent density functional theory
(TDDFT)~\cite{{20},{21},{22},{23},{24},{25}}, we explore several
previously unreported ferroic allotropes of two-dimensional (2D)
Nb$_2$O$_2$I$_4$ beyond the established ferroelectric (FE) phase.
Our simulations reveal that the electrical polarization of
Nb$_2$O$_2$I$_4$ can be reversed via anharmonic distortions of two
$\Gamma$-point phonon modes, triggered by optically induced
electronic excitation. Furthermore, activating phonon modes at the
$Y$-point facilitates the emergence of three antiferroelectric
(AFE) allotropes. By exciting an additional phonon mode at the
$\Gamma$-point, a ferrielectric (FiE) phase is induced from the
initial FE state. The optical activation of these specific modes
can be precisely controlled by modulating laser pulse parameters,
enabling the selective targeting of Nb$_2$O$_2$I$_4$ allotropes
with distinct ferroic orders. These newly predicted allotropes
exhibit energetic stability comparable to the FE phase, suggesting
their potential reversibility. Indeed, our molecular dynamics (MD)
simulations demonstrate that the FiE phase can be switched back to
the FE phase using appropriately tuned laser pulses. Additionally,
thermal activation facilitates the transition of the AFE-1 and AFE-3 phases to the FiE phase, while the AFE-2 phase reverts
directly to the FE phase. Besides, this study uncovers a feasible mechanism
for optically flipping the in-plane ferroelectric polarization of
2D Nb$_2$O$_2$I$_4$, a task that is challenging to achieve using a
conventional external electric field along the basal plane. These
findings suggest that the optical manipulation of polarized states
in 2D ferroelectric materials holds significant promise for
advanced non-binary data storage.

\section{Computational Techniques}

The real-time time-dependent density functional theory (rt-TDDFT)
calculations were performed using the time-dependent ab initio
package (TDAP)~\cite{{si-1},{si-2}}. In these rt-TDDFT
calculations, a 4$\times$2$\times$1 supercell of FE-Nb$_2$O$_2$I$_4$ was
employed, with a vacuum layer of 25~$\AA$ thickness. The Brillouin
zone was sampled using a Monkhorst-Pack scheme with a
1$\times$1$\times$1 k-point mesh for the supercell. The
time-dependent Kohn-Sham wave functions of the system evolve in a
microcanonical ensemble with a time step of 50 attoseconds.
Electron-nuclear interactions are described by Perdew-Burke-Ernzerhof
(DFT-PBE) exchange-correlation functional~\cite{PBE}. Numerical atomic
orbitals with double zeta polarization (DZP) are used as the basis
set, and the plane-wave energy cutoff was set to 200 Ry. Density
functional theory (DFT) calculations were carried out using the
SIESTA code~\cite{si-4}. Geometries were optimized using the
conjugate gradient (CG) method~\cite{si-5}, with convergence
criteria set such that no residual Hellmann-Feynman forces exceeded
10$^{-3}$~eV/$\AA$. The phonon dispersion spectra were calculated
using the Phonopy code~\cite{{si-6},{si-7}}.

\begin{figure}[t]
\includegraphics[width=0.8\columnwidth]{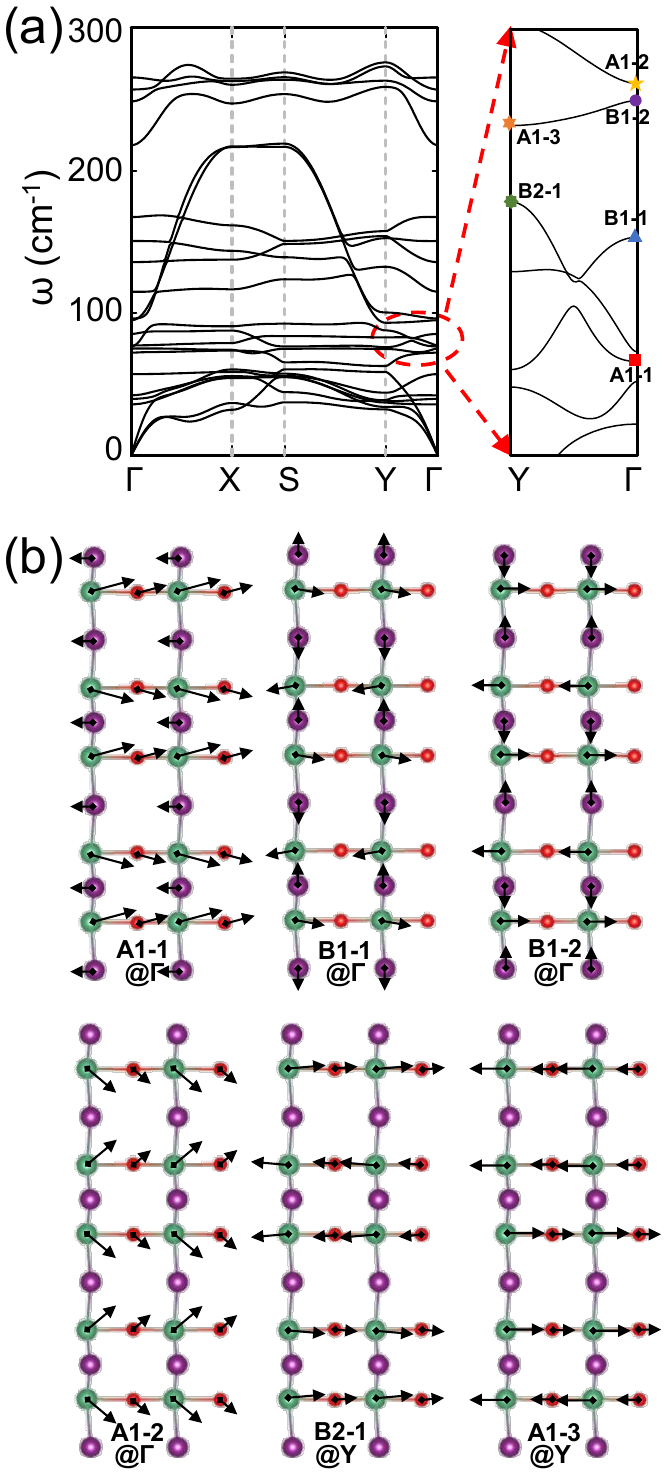}
\caption{%
(a) Phonon spectra of 2D FE-Nb$_2$O$_2$I$_4$ calculated using the
DFT-PBE energy functional. The details circled by the red dashed
line are magnified and presented in the right panel. Six special
phonon modes activated by laser pulse and determinant for the phase
transitions are highlighted by yellow pentagram, purple circle,
blue triangle, red rectangle, orange hexagram and green heptagram.
(b) The vibration eigenvectors of the six special phonon modes
of (a).%
\label{fig2}}
\end{figure}

\section{Results}

\subsection{Optically induced ferroic allotropes of Nb$_2$O$_2$I$_4$}

Nb$_2$O$_2$I$_4$ possesses a layered structure in which each layer
consists of niobium (Nb) atoms coordinated by oxygen (O) and iodine
(I) atoms. These layers are held together by weak van der Waals
forces, enabling the material to be easily exfoliated into thin
two-dimensional flakes~\cite{{vdw-2},{vdw-3},{vdw-4},{26}}. As
shown in Fig.~\ref{fig1}(a), an off-center distortion arises from
the displacement of Nb atoms from the center of the oxygen–iodine
octahedral cage along the x-direction, resulting in intrinsic
in-plane ferroelectricity with an electrical polarization of
141~pC/cm$^2$. The direction of polarization is indicated by blue
and red arrows and background colors. The atomic structure of
monolayer Nb$_2$O$_2$I$_4$ is depicted in Fig.~\ref{fig1}(b). In
this study, we demonstrate that not only can the electrical
polarization of 2D ferroelectric Nb$_2$O$_2$I$_4$ be reversed under
laser illumination, but four additional ferroic allotropes with
distinct polarization states can be obtained by tuning the laser
frequency $\omega$ and peak amplitude $E_0$, as illustrated in
Fig.~\ref{fig1}(c).The laser frequency $\omega$ used to induce
phase transitions is selected based on the electronic properties of
2D FE-Nb$_2$O$_2$I$_4$. As shown in Appendix A, monolayer
Nb$_2$O$_2$I$_4$ exhibits semiconducting behavior with a direct
bandgap of 0.85~eV. The calculated optical absorption spectrum
displays several prominent peaks; among them, we select three
photon energies of 2.1~eV, 3.1~eV, and 4.25~eV which are readily
available in the laboratory, as references for the laser frequency.

Figure~\ref{fig1}(d) presents the four ferroic phases of
Nb$_2$O$_2$I$_4$ obtained from the FE phase upon illumination with
laser pulses of different energies and electric field amplitudes.
Specifically, irradiation with a laser pulse of $\hbar\omega =
2.1$~eV and $E_0 = 0.206$~V/$\AA$ transforms the FE phase into an
antiferroelectric phase denoted as AFE-1, in which the local
electrical polarizations of the Nb-O-I cages containing Nb$_{1}$
and Nb$_{2}$ atoms are oriented to the right, whereas those
containing Nb$_{3}$ and Nb$_{4}$ atoms are oriented to the left.
Increasing the photon energy to $\hbar\omega = 3.1$~eV with $E_0 =
0.154$~V/$\AA$ yields a ferrielectric phase, in which the local
electrical polarizations of the three cages containing Nb$_{1/2/4}$
atoms align in the same direction, opposite to that of the cage
containing the Nb$_{3}$ atom. Consequently, the FiE phase exhibits
a net electrical polarization of 80.4~pC/cm$^2$.

At a higher photon energy of $\hbar\omega = 4.25$~eV, two distinct
antiferroelectric allotropes of AFE-2 and AFE-3 are obtained with
$E_0 = 0.154$~V/$\AA$ and $E_0 = 0.179$~V/$\AA$, respectively.
Although the net electrical polarization of these three AFE phases
is zero, their local polarization states differ. As shown in
Figure~\ref{fig1}(d), in the AFE-2 phase, the local electrical
polarizations of the cages containing Nb$_{1}$ and Nb$_{3}$ atoms
are oriented to the left, while those containing Nb$_{2}$ and
Nb$_{4}$ atoms are oriented to the right. While in the AFE-3 phase,
the local electrical polarizations between the closet neighboring
Nb-O-I cages along the $\vec{a}_2$ direction oriented oppositely.

These four ferroic allotropes are nearly as energetically stable as
the FE phase. Their dynamical stability is corroborated by the
calculated phonon spectra presented in Appendix B, where the
absence of imaginary frequencies confirms that all allotropes are
dynamically stable. Regarding electrical polarization and atomic
arrangement, the FE and AFE-3 phases possess nearly identical unit
cell dimensions, whereas the lattice vector $\vec{a}_2$ of the FiE,
AFE-1, and AFE-2 phases is twice that of the FE phase. Similar to
the FE phase, the remaining four ferroic phases display
semiconducting properties with direct bandgaps of approximately
0.85~eV, further details are provided in Appendix B.

\subsection{Mechanism of phase transition through activation of coherent phonon modes}

The microscopic origin of these phase transitions lies in the
specific anharmonic distortions of atoms from their equilibrium
positions. Unlike thermal activation, which may excites a broad
ensemble of phonon modes, optical excitation enables the selective
activation of coherent phonon modes. As mentioned above, the unit
cell size along the $\vec{a}_2$ direction of the AFE-1, AFE-2, and
FiE phases is twice that of the FE and AFE-3 phases.
Consequently, the wave vector connecting the $\Gamma$-point to the
high-symmetry $Y$-point at the boundary of the first Brillouin zone
(BZ) of the FE phase folds back to the $\Gamma$-point in the BZ of
the AFE-1/2 and FiE phases due to the doubled lattice periodicity.
To elucidate the contributions of distinct phonon modes to the phase
transitions originating from the FE phase, we project the atomic
displacements obtained from TDDFT-MD simulations onto the
vibrational eigenmodes at both the $\Gamma$- and $Y$-points. Six
phonon modes are identified as being predominantly excited during
the simulations. As depicted in Fig.\ref{fig2}(a), four of these
modes labeled A1-1, B1-1, B1-2, and A1-2, are located at the
$\Gamma$-point, while the remaining two modes, B1-2 and A1-3, are
associated with the $Y$-point. The vibrational eigenvectors of
these six phonon modes are illustrated in Fig.\ref{fig2}(b).

\begin{figure}[t]
\includegraphics[width=1.0\columnwidth]{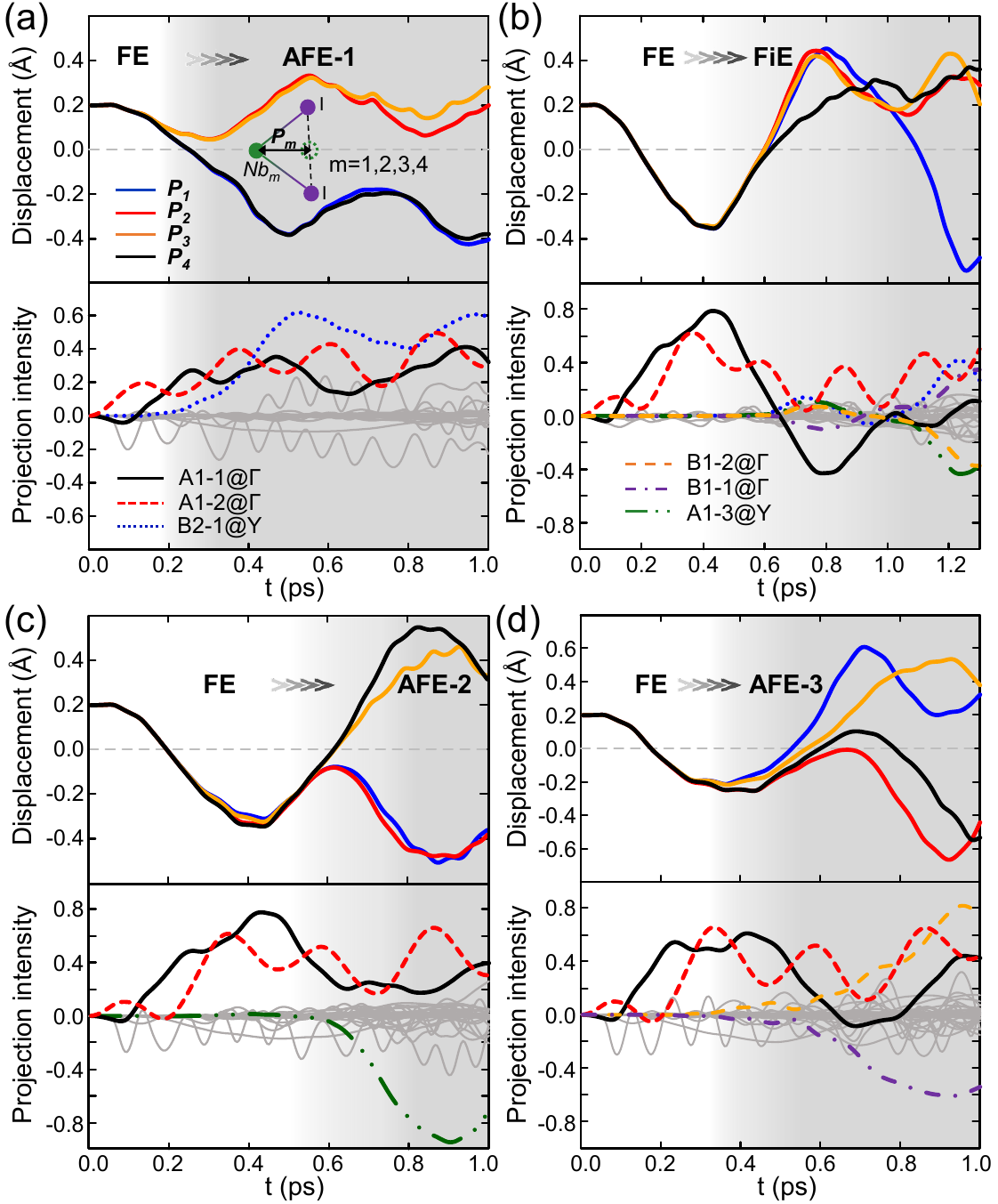}
\caption{%
The evolution of the four order parameters characterizing phase
transition of Nb$_2$O$_2$I$_4$ and the atomic displacements
projected onto phonon modes at $\Gamma$-point and $Y$-point during
the TDDFT simulations with illumination of laser pulse of (a)
$\hbar\omega$=2.1~eV, $E_0$=0.206~V/$\AA$, (b)$\hbar\omega$=3.1~eV,
$E_0$=0.154~V/$\AA$, (c)$\hbar\omega$=4.25~eV, $E_0$=0.154~V/$\AA$
and (d)$\hbar\omega$=4.25~eV, $E_0$=0.179~V/$\AA$.%
\label{fig3}}
\end{figure}

The electrical polarization is determined by the displacement of Nb
atoms from the centers of the O/I cages. As illustrated in
Fig.~\ref{fig3}(a), we define an order parameter $P_m$, where $m$
indexes the Nb atoms within the unit cell, to characterize this
displacement. The phase transition processes induced by laser
illumination are monitored via the time evolution of $P_m$.
Given that the unit cells of the AFE-1, AFE-2, and FiE phases each
contain four Nb atoms, $m$ ranges from 1 to 4. As shown in Fig.\ref{fig3}, a
consistent feature across all four laser illumination conditions is
the initial activation of the $A1-1$ and $A1-2$ phonon modes during
the TDDFT-MD simulations. As indicated in Fig.\ref{fig2}(b), the
vibrational eigenvectors of both modes exhibit the same phase,
driving correlated motions between all four Nb atoms and their
surrounding I atoms. While these collective displacements facilitate
a reversal of the in-plane polarization in, they
do not trigger a full transition to the FE-Nb$_2$O$_2$I$_4$. This
mechanism is corroborated by the nearly synchronous decrease of all
$P_m$ values during the initial period of each phase transition, as
evidenced in Fig.\ref{fig3}.

As time evolves, the activation of phonon modes under the four
different laser illumination conditions also varies. For the laser
pulse with $\hbar\omega$=2.1~eV and $E_{0} = 0.206V/{\AA}$, before
the polarization reversal of FE-Nb$_2$O$_2$I$_4$ is completed, an
additional phonon mode of $B2-1$ at the $Y$-point is excited at
approximately 0.2~ps, as shown in Fig.~\ref{fig3}(a). As
illustrated in Fig.~\ref{fig2}(b), this mode induces opposite
displacements of the Nb$_{1/4}$ and Nb$_{2/3}$ atoms. The
superposition of the anharmonic motions of this phonon modes
causes $P_{2/3}$ to remain positive while $P_{1/4}$ becomes
negative, leading to the transformation from the FE phase to the
AFE-1 phase at $\sim 0.22$~ps. The system then remains in the
AFE-1 phase for approximately 0.8~ps until the end of our TDDFT-MD
simulation. From Fig.~\ref{fig3}(a), we note that other phonon
modes are also activated during the process. To confirm that these
three specific modes of $B2-1$ is the key
determinants of the AFE-1 phase formation and that no other
critical mode has been overlooked, we present the structural
evolution of the FE phase in Appendix C, obtained by superimposing
only this mode. The resulting evolutions of $P_m$
qualitatively match those in Fig.~\ref{fig3}(a), and the
AFE-1 phase is again formed at approximately 0.4~ps. This confirms
that $B2-1$ is indeed the decisive factors
driving the transition.

Under illumination of a laser pulse with $\hbar\omega = 3.1$~eV and
$E_0 = 0.154$~V/$\AA$, only the $A1-1$ and $A1-2$ phonon modes are
activated during the initial 0.6~ps, as shown in
Fig.~\ref{fig3}(b). Compared to the case with $\hbar\omega =
2.1$~eV, the activation amplitudes of these two modes are
approximately twice as large, indicating an acceleration of the
polarization reversal process induced by them. From
Fig.~\ref{fig3}(b), we observe that the 0.6~ps duration is
sufficient for the anharmonic oscillations of these two modes to
reverse the polarization in the FE-Nb$_2$O$_2$I$_4$. Specifically, the order
parameter $P_m$ changes from positive to negative at $t =
0.2$~ps. At $t = 0.6$~ps, other four additional phonon modes of
$B2-1$, $B1-2$, $B1-1$, and $A1-3$ become activated, albeit with
small amplitudes. The atomic displacements of Nb$_2$O$_2$I$_4$ remain
dominated by the $A1-1$ and $A1-2$ modes, causing $P_m$ to switch
from negative back to positive, thus, the electrical polarization
reverses again. As time evolves to $t = 1.0$~ps, the activation
amplitudes of $B2-1$, $B1-2$, $B1-1$, and $A1-3$ increase and
become comparable to those of $A1-1$ and $A1-2$. Under the combined
effect of these four modes, $P_1$ decreases rapidly and becomes negative
at $t = 1.1$~ps, while the other three order parameters remain
positive. This behavior signifies the transformation of FE phase
into the FiE phase. To further validate the role of these modes, we
studied the atomic evolution of the FE phase by superposing these
four modes in Appendix C. The resulting
evolutions of $P_m$ qualitatively match those presented in
Fig.~\ref{fig3}(b), confirming that these four phonon modes are
indeed the decisive factors driving the phase transition and that
no other non-negligible modes have been omitted.

\begin{figure}[t]
\includegraphics[width=1.0\columnwidth]{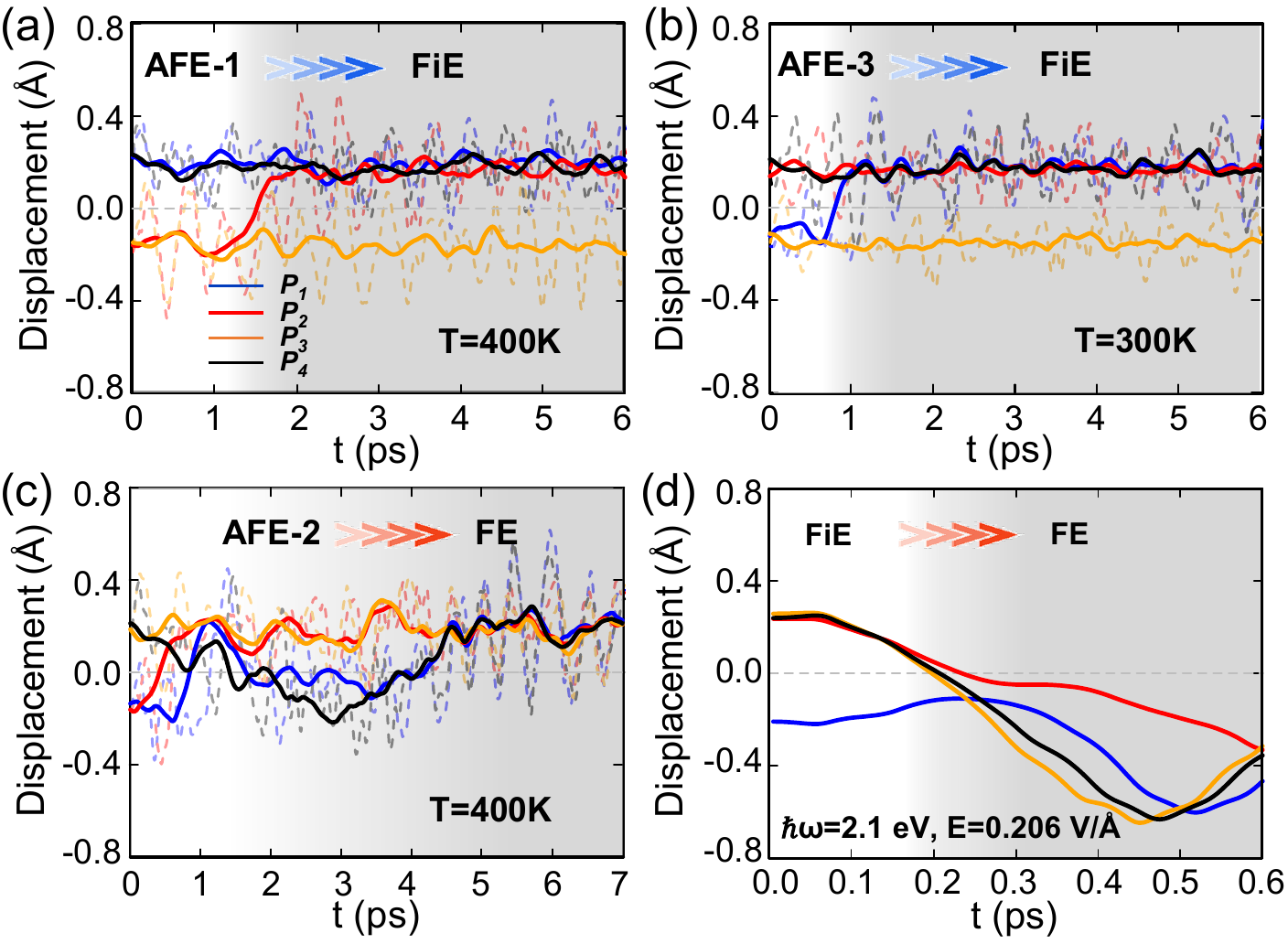}
\caption{%
The evolution of the order parameters during the NVT-MD simulation
of (a) AFE-1 phase with T=400~K, (b) AFE-3 phase with T=300~K, (c)
AFE-2 phase with T=400~K. (d) The evolution of the order parameters
of FiE phase during TDDFT simulation under illumination of laser
pulse with $\hbar\omega$=2.1~eV, $E_0$=0.206~V/$\AA$.%
\label{fig4}}
\end{figure}

Increasing the photon energy of the laser pulse to $\hbar\omega
= 4.25$~eV and applying two different electric fields, $E_0 =
0.154$~V/$\AA$ and $E_0 = 0.179$~V/$\AA$, results in two distinct
phase transitions, as shown in Fig.~\ref{fig3}(c) and (d). In both
cases, phonon modes of $A1-1$ and $A1-2$ are activated at early
time period, with activation magnitudes large enough to induce
reversal of the electric polarization, as evidenced by the change
in $P_m$ from positive to negative. For the case with $E_0 =
0.154$~V/$\AA$ , an additional phonon mode of $A1-3$ is activated
at $\sim 0.6$~ps with a magnitude larger than those of $A1-1$
and $A1-2$, as shown in Fig.~\ref{fig3}(c). Due to the anharmonic
vibration of $A1-3$, $P_{1/2}$ and $P_{3/4}$ evolve in opposite
directions. Consequently, the FE phase transitions to the AFE-2
phase. For the other case with $E_0 = 0.179$~V/$\AA$, two
additional phonon modes, $B1-1$ and $B1-2$, are activated at
$\sim 0.4$~ps. Their activation magnitudes increase and become
comparable to those of $A1-1$ and $A1-2$ by $\sim 0.7$~ps.
After this moment, $P_{1/3}$ and $P_{2/4}$ evolve in opposite
directions, and the AFE-3 phase is formed by $\sim 0.8$~ps, as shown
in Fig.~\ref{fig3}(d). To verify the uniqueness of these phonon
modes in driving the phase transition, we separately studied the
atomic evolution of the FE phase when superposed exclusively with
these modes under each of the two laser illuminations, which is
presented in Appendix C. The qualitative fitting of the evolution
of $P_m$ indicates that these phonon modes are indeed the key
factors. Additional movies about the TDDFT-MD simulations to
illustrate the above phase transitions from FE phase to the other
ferroic phases are presented in the Appendix D.

\subsection{Reversibility of the phase transitions}

For Nb$_2$O$_2$I$_4$ to be a viable candidate for
F-RAM applications, a critical prerequisite is the reversibility of
the transitions between its ferroic phases. As previously noted,
these phases are almost energetically degenerate, implying that
transitions from the FE phase can potentially be reversed. Rather
than employing the nudged elastic band (NEB) method to investigate
static transformation pathways, we adopt more realistic molecular
dynamics approaches incorporating thermal and optical activation to
elucidate the underlying transformation mechanisms.

Unlike the FE phase, which readily transforms into other ferroic
phases under laser illumination, the three AFE phases have not been
observed to revert to the FE phase via optical activation. However,
as illustrated in Fig.~\ref{fig4}(a) and (b), both the AFE-1 and AFE-3
phases can transition to the FiE phase through thermal
activation. To mitigate the effect of order parameter oscillations
that may obscure intrinsic physical behavior in MD simulations, we
employ a simple moving average (SMA) method to qualitatively
analyze their evolution. Our NVT-MD simulation with $T = 400$~K
shown in Fig.~\ref{fig4}(a), reveals that the signs of the order
parameters $P_{1}$, $P_{3}$, and $P_{4}$ of Nb$_2$O$_2$I$_4$ remain
unchanged, while $P_{2}$ switches from negative to positive at
$\sim$1.5~ps, indicating a transition from the AFE-1 phase to the FiE
phase. Similarly, at temperature of $T = 300$~K, $P_{1}$ undergos a
sign change around $\sim$1.0~ps as shown in Fig.~\ref{fig4}(b),
suggesting a transition from the AFE-3 phase to the FiE phase.
Under thermal activation at $T = 400$~K, the AFE-2 phase exhibits
significant fluctuations in three order parameters of $P_{1}$,
$P_{2}$ and $P_{4}$, as shown in Fig.~\ref{fig4}(c).
At $\sim$4.0~ps, all of the order parameters
evolve to be positive and the system transformed from the FiE phase
to the FE phase which remains stable until the end of our MD
simulation.

As illustrated in Fig.~\ref{fig4}(a) and (b), the AFE-1 and AFE-3
phases transition to the FiE phase under thermal activation, and
the FiE phase remains stable within the time limitation of our
calculations, suggesting that the FiE phase does not readily revert
to other phases under thermal excitation. However, as shown in
Fig.~\ref{fig4}(d), the same laser pulse used to induce the phase
transition from FE phase to FiE phase can also trigger the reverse process.
Consequently, the AFE-1 and AFE-3 phases can be
indirectly transformed back into the FE phase via the intermediate FiE
phase. All phase transitions are supported by corresponding moives
of MD simulations, available in Video~\ref{Video1} of Appendix D.

\begin{figure*}[t]
\includegraphics[width=1.8\columnwidth]{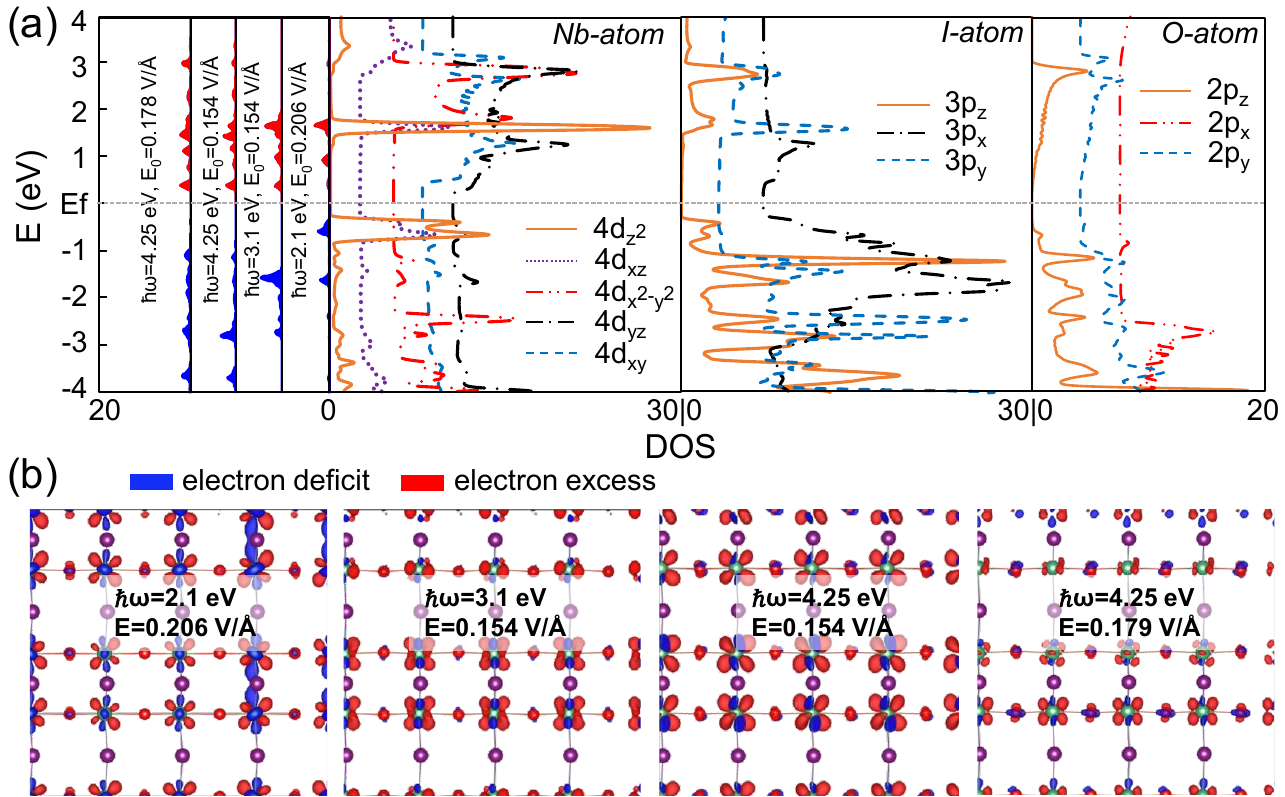}
\caption{%
(a) Energy distribution of the excited electrons (red area) and the
holes (blue area) with the illumination of the four laser
pulses. Electronic density of states is projected on 4d-orbitals of
the Nb atoms, 3p-orbitals of the I atoms and 2p-orbitals of O
atoms. (b) The corresponding charge redistribution before and after
the photo excitations, we display the difference charge density
$\delta\rho$=$\rho(t=105~fs)-\rho(t=0~fs)$ as isosurfaces bounding
regions of electron excess at $+3.5{\times}10^{-3}~\text{e}$/{\AA}$^3$, shown
in red, and electron deficiency at $-3.5{\times}10^{-3}~\text{e}$/{\AA}$^3$,
shown in blue.%
\label{fig5}}
\end{figure*}

\section{Discussion}

In the main text, we demonstrate a direct correlation between the
activation of specific phonon modes and the observed phase
transitions. Usually, during the TDDFT-MD simulations, the
optically modified potential energy surface differs from the
adiabatic case~\cite{{17},{lpw}}, suggesting that optically controlled phase
transitions may be induced by nonadiabatic coupling(i.e.,
correlated electron–phonon dynamics beyond the Born–Oppenheimer
approximation). It is therefore necessary to investigate the
changes in the electronic structure under laser illumination. As
shown in Fig.~\ref{fig5}(a), we present the electron excitation
upon termination of the laser pulses. With increasing photon energy
$\hbar\omega$, electrons from deeper bands below the Fermi level
are excited to higher bands above the fermi level. For the same
photon energy $\hbar\omega = 4.25$~eV but different electric field
strengths $E_0$, the electronic excitations are qualitatively
similar but quantitatively different. In the case with higher $E_0$
approximately 0.125 more electrons per unit cell are excited. A
more direct illustration of the electron excitations is provided in
Fig.~\ref{fig5}(b), where the optically induced charge
redistribution are similar for the four laser illuminations. This
finding aligns with the similar atomic evolution of decreasing of
order parameters at the early stage of TDDFT-MD simulations in
Fig.~\ref{fig3}, which indicating the intrinsic electron-phonon
coupling.

A key advantage of FE materials in electronic applications is the
switchability of their electrical polarization. However, achieving
polarization reversal remains a significant experimental challenge.
This is particularly true for two-dimensional FE materials with
in-plane polarization, which require a lateral electric field to
reverse the polarization. At the nanoscale, obtaining a uniform
field distribution via electrode contacts is difficult due to edge
effects and fabrication constraints. In this study, the distinct
ferroic phases arise from differences in the local polarization of
the Nb–O–I cage, characterized by the displacement of Nb atoms.
Reversing the local polarization requires an atomic shift of 0.33~$\AA$
and overcoming an energy barrier of approximately 25~meV. The
electric field needed to drive this Nb displacement is about
19~V/nm which is technically challenging to apply experimentally
and risks damaging the sample. Here, we demonstrate that in-plane
electrical polarization reversal in Nb$_2$O$_2$I$_4$ can be
achieved through optical activation of coherent phonon modes. We
anticipate that this approach may be extended to other systems
beyond the specific case of Nb$_2$O$_2$I$_4$, such as the recently
discovered sliding ferroelectricity in 2D materials, where
polarization reversal is governed by the relative sliding of atomic
layers~\cite{{huayi-1},{huayi-2},{huayi-3}}. In such systems, the response of polarization to an
external electric field or strain can be significantly delayed due
to non-negligible interlayer interactions. Optical excitation
offers an alternative route to modify the potential energy surface,
thus lowering the energy barrier and enabling efficient
polarization reversal.

\section{Summary and Conclusions}

In summary, using rt-TDDFT simulations, we observed diverse phase
transitions in the 2D FE-Nb$_2$O$_2$I$_4$. By adjusting the
parameters of the laser pulses, we selectively activate different
phonon modes. The resultant non-equilibrium atomic dynamics driven
by these phonon modes result not only in the reversal of the
in-plane polarization but also in the formation of previously
unexplored AFE and FiE phases. The similar atomic evolution at
early times and the analogous charge redistribution under different
laser pulse illuminations indicate an intrinsic electron–phonon
coupling in the system. We also find that these optically induced
ferroic phases can be transferred back to the original FE phase
either directly or indirectly. While the AFE-2 phase can be
reverted to the FE phase simply by heating the system to 400 K, the
AFE-1 and AFE-3 phases first transform into the FiE phase upon
thermal activation, this FiE phase can then be optically switched
back to the FE phase. These observed phenomena making the 2D
Nb$_2$O$_2$I$_4$ as a promising material for non-binary data
storage in future electronic applications.

\renewcommand\thesubsection{\Alph{subsection}}
\renewcommand{\theequation}{A\arabic{equation}}
\renewcommand{\thevideo}{A\arabic{video}}
\setcounter{subsection}{0} %
\setcounter{equation}{0} %

\section*{Appendix}

\begin{figure}[t]
\includegraphics[width=1.0\columnwidth]{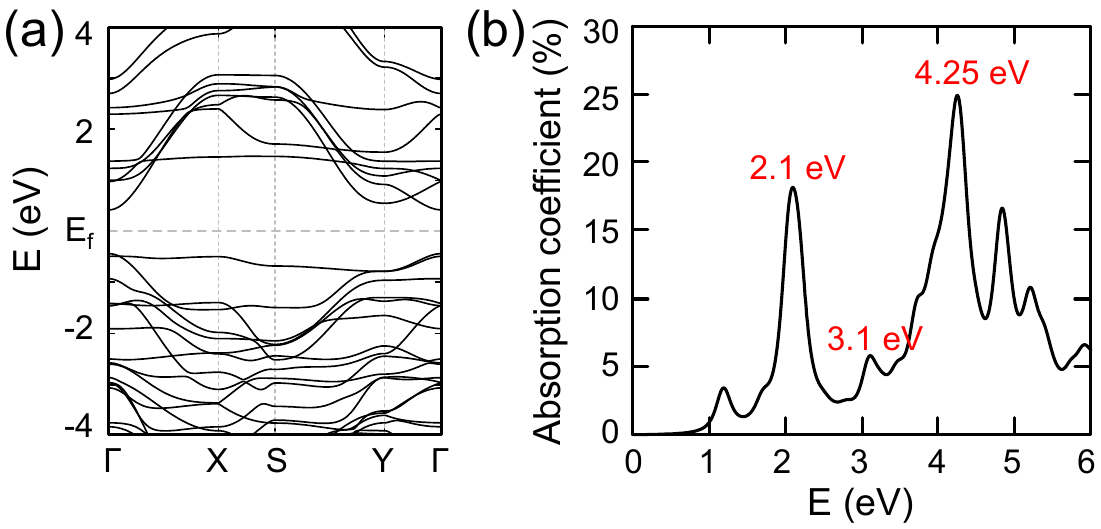}
\caption{%
(a) DFT-PBE calculated electronic band structure and (b) optical
absorption coefficient of FE-Nb$_2$O$_2$I$_4$.%
\label{fig6}}
\end{figure}

\begin{figure}[t]
\includegraphics[width=1.0\columnwidth]{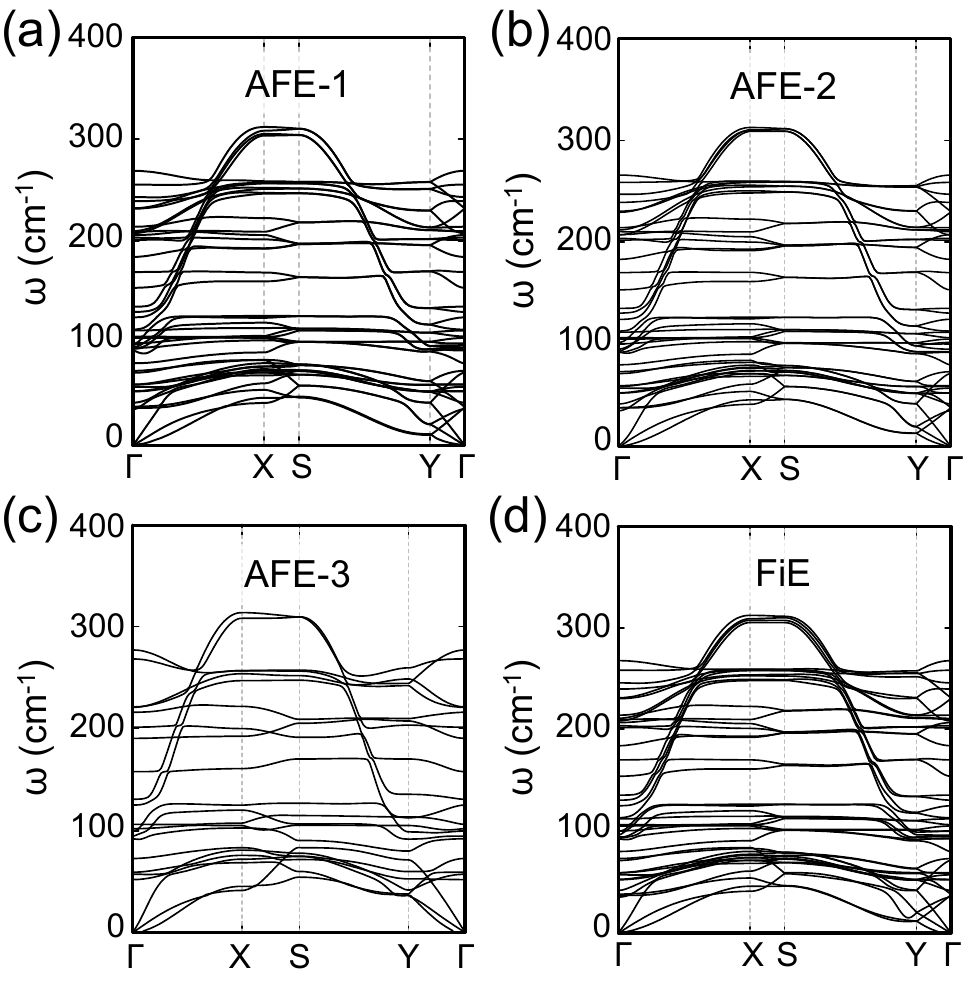}
\caption{%
DFT-PBE calculated phonon spectra of Nb$_2$O$_2$I$_4$ with (a)
AFE-1 phase, (b) AFE-2 phase, (c) AFE-3 phase and (d) FiE phase. %
\label{fig7}}
\end{figure}

\subsection{Electronic band structure and optical absorption of Nb$_2$O$_2$I$_4$ }

The DFT-calculated electronic band structure, presented in
Fig.\ref{fig6}(a), reveals the semiconducting nature of
Nb$_2$O$_2$I$_4$ with a direct bandgap of 0.88eV. Both the valence
band maximum (VBM) and the conduction band minimum (CBM) are
located at the $\Gamma$-point. To determine the appropriate laser
parameters chosen to do the TDDFT simulations, we calculated the
optical absorption coefficient of Nb$_2$O$_2$I$_4$. As shown in
Fig.\ref{fig6}(b), the spectrum exhibits six absorption peaks. We
selected the three peaks at 2.1eV, 3.1eV, and 4.25eV, which are
available at the experiment, to set the laser parameters for
investigating the optical phase transitions in Nb$_2$O$_2$I$_4$.

\subsection{Stability and electronic properties of Nb$_2$O$_2$I$_4$ }

In addition to the FE phase of Nb$_2$O$_2$I$_4$, we
calculated the phonon spectra for the other four ferroic phases, as
shown in Fig.~\ref{fig7}. No imaginary frequencies were observed,
confirming that these ferroic phases are dynamically stable.

\begin{figure}[h]
\includegraphics[width=1.0\columnwidth]{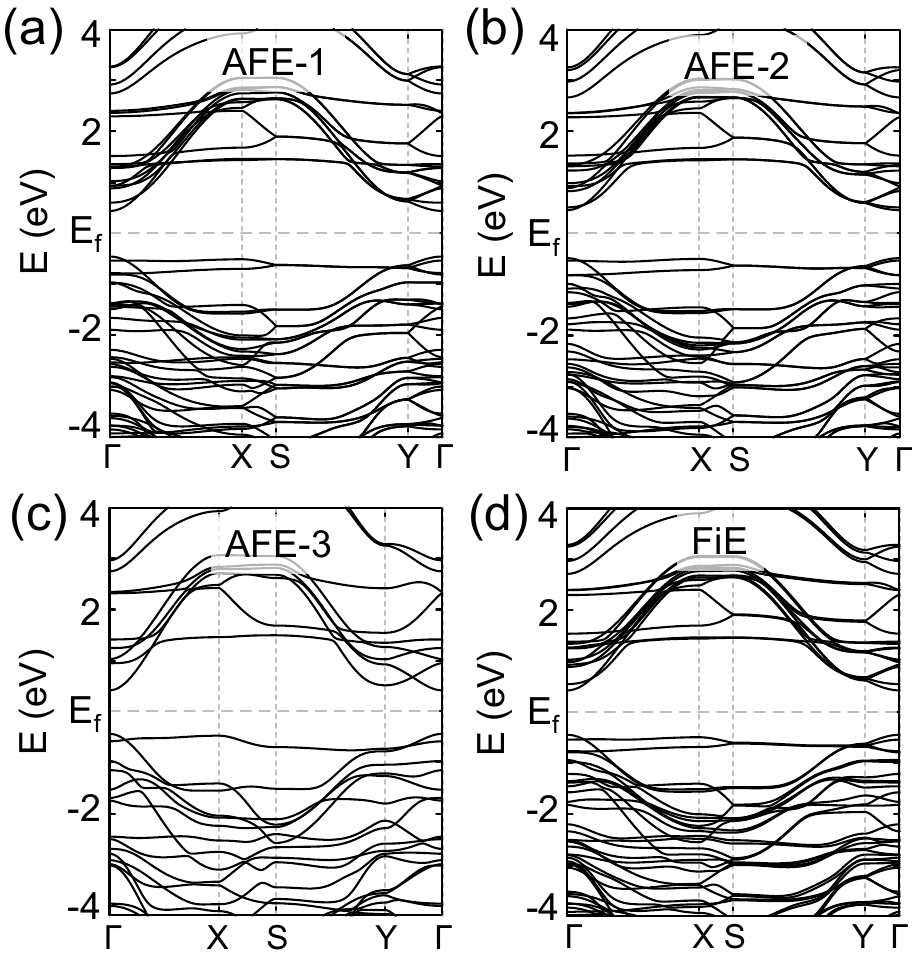}
\caption{%
Electronic band structure of Nb$_2$O$_2$I$_4$ with (a) AFE-1 phase,
(b) AFE-2 phase, (c) AFE-3 phase and (d) FiE phase.%
\label{fig8}}
\end{figure}

Similar to the FE phase, the AFE and FiE phases of Nb$_2$O$_2$I$_4$
also exhibit the characteristics of direct-bandgap semiconductors,
with both the VBM and CBM located at the $\Gamma$-point, as
illustrated in Fig.~\ref{fig8}.

\begin{figure}[t]
\includegraphics[width=1.0\columnwidth]{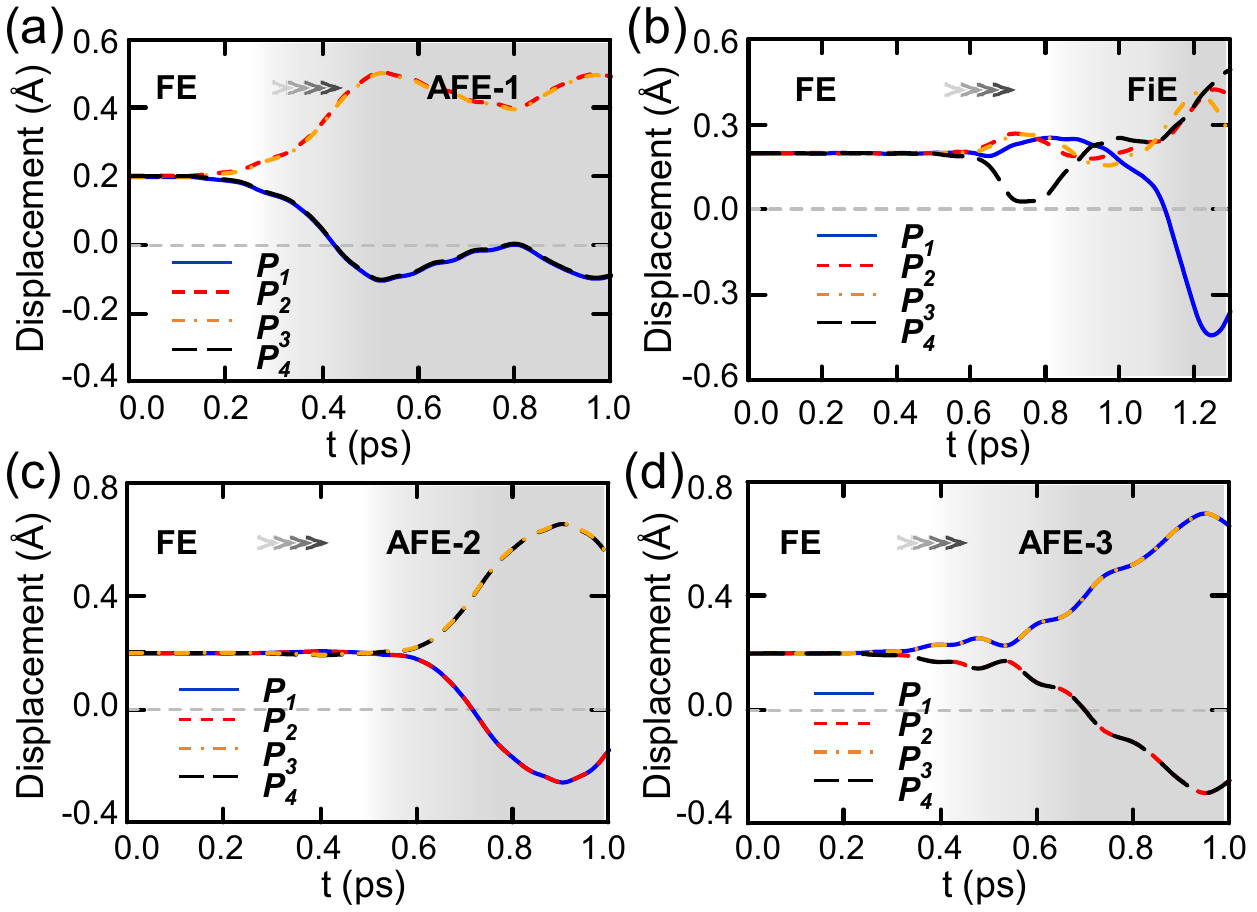}
\caption{%
The evolution of order parameters starting from the ferroelectric
phase of Nb$_2$O$_2$I$_4$ with the superposition of phonon mode of
(a) B2-1, (b) A1-3, B1-1, B1-2 and B2-1, (c) A1-3, (d) B1-1 and B1-2.%
\label{fig9}}
\end{figure}

\subsection{Unique effect of the selected phonon modes on the phase
            transitions}

For each illumination of laser pulse in Fig.~\ref{fig3}, besides
the six special phonon modes of A1-1/2/3, B1-1/2 and B2-1 are
activated, other phonon modes are also activated, but with a
relatively weak intensity. As illustrated in Fig.\ref{fig2}(b), the
atomic distortion induced by the A1-1 and A1-2 modes leads to
polarization reversal but does not result in phase transitions to
other ferroic phases. To verify whether the anharmonic distortions
induced by the other four phonon modes are the determining factors
in the phase transitions, we superimposed the atomic vibrations of
these specific modes onto the ferroelectric phase, applying the
same weights as those in Fig.\ref{fig3} and examined the atomic
evolution. Fig.\ref{fig8} displays the time-dependent change in
the order parameters. The observed evolution is qualitatively
similar to that in Fig.~\ref{fig3}, and the FE phase successfully
transforms into the targeted ferroic phases. This confirms that
A1-3, B1-1, B1-2 and B2-1 are indeed the decisive phonon modes driving
the phase transitions, and that no other critical factors were
overlooked.

\subsection{Molecular Dynamic simulations of the phase transitions}

In addition to the evolution of order parameters shown in
Figs.\ref{fig3} and \ref{fig4}, which indicate the occurrence of
phase transitions during the TDDFT and NVT-MD simulations, we
provide Video\ref{Video1} as direct evidence of these structural
transformations. Specifically, Videos~\ref{Video1}(a)–(d) depict
the phase transitions from the FE phase to the other ferroic
phases. Videos~\ref{Video1}(e) and (f) showcase the transformation
of the AFE-1 and AFE-3 phases into the FiE phase, while
Videos~\ref{Video1}(g) and (h) illustrate the reversion of the FiE
and AFE-2 phases back to the FE phase.

\begin{video}[h]
\includegraphics[width=1.0\columnwidth]{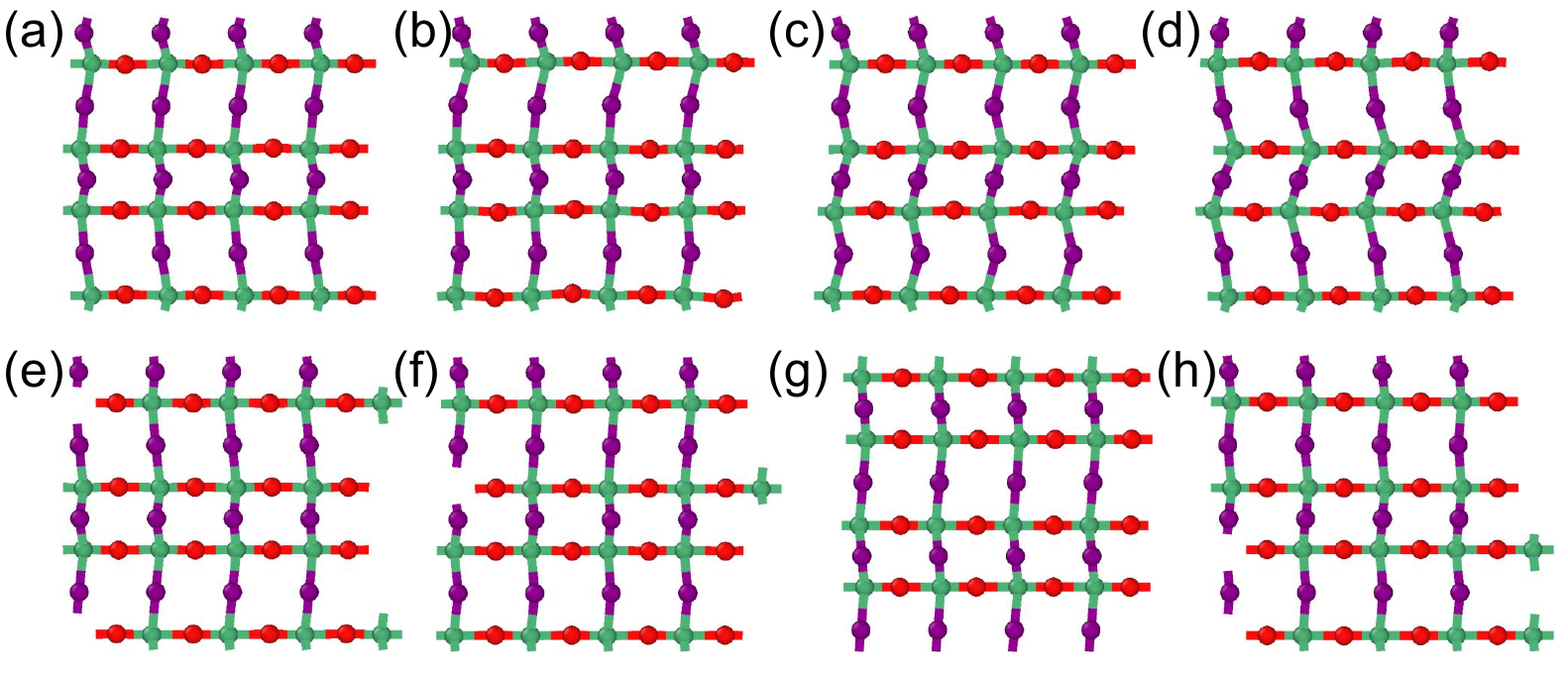}
\setfloatlink{video1.mp4,video2.mp4,video3.mp4,video4.mp4,video5.mp4,video6.mp4,video7.mp4,video8.mp4} %
\caption{
Results of TDDFT-MD simulation of phase transition from FE phase to
(a) AFE-1 phase, (b) FiE phase, (c) AFE-2 phase, (d) AFE-3 phase.
NVT-MD simulation of phase transition from (e) AFE-1 phase to FiE
phase at $T=400$~K, (f) AFE-3 phase to FiE phase at $T=300$~K, (g)
AFE-2 phase to FE phase at $T=400$~K. (h) And TDDFT-MD simulation
of phase transition from FiE phase to FE phase.%
\label{Video1}}
\end{video}

\bigskip
\begin{acknowledgements}
This study is supported by the
National Natural Science Foundation of China (Grant Nos. 62274028,
12204095), National Key Research and Development Program of China
(Grant No. 2024YFA1408603), CAS Project for Young Scientists in
Basic Research (Grant No. YSBR-120).

Chuanlin Liu and Dan Liu contributed equally to this work.
\end{acknowledgements}


%
%
%


%

\end{document}